\documentclass[onecolumn,noshowpacs,superscriptaddress,nobibnotes,nofootinbib,12pt]{revtex4}
\usepackage{amsmath,amssymb}
\usepackage{graphicx}

\begin{document}
\title{On dissociation of heavy mesons in a hot quark-gluon
plasma}
\author{Fabio Dominguez}
\email{fabio@phys.columbia.edu} \affiliation{Department of Physics,
Columbia University, New York, NY, 10027, USA}

\author{Bin Wu\footnote{B.W. is supported by China Scholarship Council.}}
\email{bw2246@columbia.edu} \affiliation{Department of Physics,
Peking University, Beijing, 100871, P.R. China}
\affiliation{Department of Physics, Columbia University, New York,
NY, 10027, USA}

\begin{abstract}
We compare two mechanisms for the dissociation of heavy mesons in an
infinite quark-gluon plasma: dynamic Debye screening and multiple
scattering. Using the uncertainty principle inspired
by a Schr\"{o}dinger-like equation, we find that the criterion
$a_B\simeq1/\mu\simeq\frac{1}{\alpha_{eff}^{1/2}T}$ with
$\alpha_{eff}\equiv\alpha(N_c+{N_f\over2})$ is parametrically
true both for the dissociation of fast moving heavy mesons with a size
$a_B$ due to dynamic Debye screening as well as for mesons at rest in the medium. In contrast, we find that the
criterion for the dissociation of heavy mesons due to uncorrelated
multiple scattering is parametrically $a_B\simeq {1\over
[\gamma\alpha_{eff}\ln{1\over \alpha_{eff}}]^{1\over 3}T}$.
Therefore, multiple scattering is a more efficient mechanism for the
dissociation of heavy mesons in an infinite hot plasma.
\end{abstract}

\date{\today}
\maketitle
\newpage

\section{Introduction}
The problem of dissociation of bound states in a hot QCD medium is of great importance in heavy ion collisions as it provides evidence for the creation of the quark-gluon plasma in heavy ion collisions \cite{Leitch:2008vw}. In order to get a better understanding of the properties of this state of matter it is necessary to establish criteria under which bound states are not allowed to exist or are broken apart. We focus on the study of two particular mechanisms which we believe to be the main causes of dissociation in the plasma: Debye screening and multiple scattering with constituents of the plasma.

In Ref.\cite{Matsui:1986dk}, the authors discussed $J/\psi$
suppression due to Debye screening by the quark-gluon plasma and
the importance of this signature to diagnose quark-gluon plasma
formation in heavy ion collisions. They found that the criterion
for the dissociation of a $J/\psi$ at rest is $r_{J/\psi}=1.61 r_D$
with $r_D\equiv\frac{1}{\sqrt{2}\mu}$, the Debye screening length. In contrast, for a heavy quark-antiquark pair
moving at velocity $v$ in an infinite strongly coupled $\mathcal{N}=4$ SYM
plasma, the AdS/CFT calculation shows that the screening length goes like
$r_D(v,T)\sim r_D(0,T)/\sqrt{\gamma}$ \cite{Liu:2006nn}. This suggests we may expect a
criterion for the dissociation of heavy quarkonia in such a strongly
coupled plasma of the form $r_{Q\bar{Q}}\sim r_D(0,T)/\sqrt{\gamma}$. As for the dissociation due to multiple scattering with the constituents of the plasma, Ref.
\cite{Adil:2006ra} addresses the problem of heavy meson suppression
in a finite dense QCD medium and predicts suppression of $B$-mesons
comparable to that of $D$-mesons at transverse momenta as low as
$p_T\sim 10$ GeV.

The calculation of the screening effect for mesons at rest is not enough to understand the suppression observed in the data from heavy ion collisions. It is also important to establish when a fast moving meson is broken apart due to the presence of the plasma. The latter case is the one explicitly addressed in this paper. For the screening effect, the full calculation is too complicated to be performed analytically and there is not an obvious way of getting a simple picture where we can get a good estimate based on the uncertainty principle as done in \cite{Matsui:1986dk}. By considering the Dirac equation in light-cone coordinates and in light-cone gauge we propose a Schr\"odinger-like equation where we can rely on the uncertainty principle to get a sensible estimate for the criterion for dissociation of bound states. This approach requires the calculation of the effective field produced by a fast moving charge with respect to the plasma, which is done in the appropriate region of momenta where we keep terms only up to first order in ${k_\perp\over k_z}\sim\gamma^{-1}$. Even though in covariant gauge the effective
field in the rest frame of the charge moving relative to the plasma
with $v\simeq1$ is highly anisotropic\cite{Chu:1988wh}, we find that
after a gauge transformation and going back to the rest frame of
the plasma, the anisotropy is suppressed as inverse powers of
$\gamma$ compared to that due to the Lorentz contraction. We find
that the criterion $a_B\simeq{1\over\mu}$ is also parametrically
true for fast moving heavy mesons.

For multiple scattering we get a criterion for dissociation in terms of the saturation momentum of the system $Q_{s}$, namely $a_{B}\simeq 1/Q_{s}$. For this purpose we first identify the typical time between interactions inside the meson $\tau_{B}$. Then the criterion for dissociation is given by the statement that if the quarks inside the meson pick up enough transverse momentum during that period then the meson breaks up. We take as the natural scale for transverse momentum inside of the heavy meson $1/a_{B}$. The transverse momentum broadening is given by $Q_{s}$ where the role of the lenght of the plasma is played by $\tau_{B}$ (assuming an infinite plasma). In the case of uncorrelated multiple scattering, the criterion $Q_s\simeq 1/a_B$ gives
\begin{equation}
a_B\simeq {1\over [\gamma\alpha_{eff}\ln{1\over
\alpha_{eff}}]^{1\over 3}T},
\end{equation}
with $\alpha_{eff}\equiv\alpha(N_c+{N_f\over2})$. It is also
parametrically true for the dissociation of heavy mesons almost at
rest with the plasma. Comparing this result with the criterion obtained from the screening effect we conclude that, in an infinite plasma, multiple scattering is a more efficient mechanism for the dissociation of heavy mesons.

The paper is organized as follows. In Sec. II, by analyzing the
Dirac equation, we conclude that those photons with
$|k_\perp|\lesssim 1/a_B$ and $|k_z|\lesssim \gamma/a_B$ are
essential for the binding in the partonic language and get a
Schr\"{o}dinger-like equation in light-cone gauge in light-cone
coordinates. We also discuss the typical time scale
$\tau_B=\gamma/E_B$ in a heavy meson. In Sec. III, we calculate the
effective field induced by a fast moving charge in light-cone gauge
up to the first order in ${k_\perp\over k_z}\sim\gamma^{-1}$ and use
the uncertainty principle to estimate the criterion for the
dissociation of fast moving heavy mesons. In Sec. IV, we give a
parametric estimate for the dissociation of
heavy mesons due to multiple scattering in an infinite hot
quark-gluon plasma. In the Appendix, we compare the classical field
$A^\mu$ calculated from classical electrodynamics and from QFT to
illustrate the connection between the classical field and virtual
photons.

\section{The Dirac Equation for a fast moving bound state}

Even though the correct mathematical treatment of a relativistic
two-body system can be done by means of the Bethe-Salpeter
equation, we choose to use the Dirac equation instead in order to
get a simpler physical picture. This approximation is valid when one
of the particles involved is much heavier than the other, in which
case the field produced by the heavier particle is no longer a
dynamical variable and can be treated as an external field. In this
framework, we can investigate the general properties of the wave
function of the lighter quark and determine under which conditions
it will be bound. In order to be able to treat the system
perturbatively we still have to assume both masses are much larger
than $\Lambda_{QCD}$ with one of the masses much greater than the
other. Although the formal results are only valid in this case, we
believe the parametric result should be the same for the case of
equal masses. We assume that the plasma is in the deconfinement
phase and in the perturbative regime. In this circumstance, we can
treat the problem in the quark-gluon plasma in a similar way as that
in a QED plasma.

In the following we first investigate the role played
by photons with momenta in different regions in the binding of a
fast moving bound state and address the question of what is the
approximate equation, appropriate for such a system, analogous to
the Schr\"{o}dinger equation for a bound state at rest. Then, we
answer the question under which circumstances our analysis of the
QED bound states applies to heavy mesons. We also give a brief
illustration about the typical time scale $\tau_B=\gamma/E_B$ based
on the perturbative definition of the wave function at the end of
this section.

Since the detailed screening effect for a moving bound state is too
complicated to be solved analytically \cite{Chu:1988wh}, we try to
simplify the problem and make it suitable for an intuitive
understanding by analyzing the Dirac equation for a fast moving
bound state and determining which photons are essential for the
binding. This will allow us to make appropriate approximations to
the calculation of the effective field induced by the heavier
particle in the presence of the plasma. With the effective field, we
still need a Schr\"{o}dinger-like equation in light-cone coordinates
which manifests the uncertainty principle and simplifies the spin
structure in order to enable us to use a physical analysis of fast
moving bound states similar to that for bound states at rest in Ref.
\cite{Matsui:1986dk}. In the following analysis, we assume that the
lighter particle has a mass $m_a$ and a charge $e_a$ while the
heavier particle has a mass $m_b$ and a charge $e_b$. The whole
analysis in the following applies to heavy mesons provided the
quarks are heavy enough.

\subsection{Which photons are responsible for the binding?}

In the vacuum case the field $A^\mu$ is much simpler, in certain
gauges, in the rest frame of the bound state. Therefore, Let us start
with this simpler case and assume that we have already solved
$\Psi_{0n}(p)$, the wave function in the rest frame of the bound
state, which peaks at $\vec{p}=0$ with a width $\Delta p=1/a_B$, and
$p^0=E_n$. By boosting to the lab frame, we obtain $\Psi_{n}(p)$
which peaks at $\vec{p}_{0\perp}=0$ with $\Delta p_\perp=1/a_B$,
$p_{0z}=v\gamma E_n$ with $\Delta p_z=\gamma/a_B$ and $p_0^0=\gamma
E_n$. By writing
\begin{equation}
\Psi_{n}(p)=\left(
        \begin{array}{c}
            \varphi(p)\\
            \chi(p)
        \end{array}
        \right),\label{WaveFunctionD_s}
\end{equation}
and inserting it into the Dirac equation in momentum space
\begin{equation}
\begin{split}
\left(p\llap{$/$}-m_a\right)\Psi_{n}(p)&=e_a\int {d^4k\over
(2\pi)^4}\left[A\llap{$/$}(k) \Psi_n(p-k)\right],\label{de}
\end{split}
\end{equation}
we get, in the chiral representation\cite{Peskin:1995ev},
\begin{equation}
        \left(
        \begin{array}{c}
            p\cdot\sigma\chi(p)-m_a\varphi(p)\\
            p\cdot\bar{\sigma}\varphi(p)-m_a\chi(p)
        \end{array}
        \right)
        =e_a\int{d^4k\over (2\pi)^4}\left(
        \begin{array}{c}
            A(k)\cdot\sigma\chi(p-k)\\
            A(k)\cdot\bar{\sigma}\varphi(p-k)
        \end{array}
        \right).\label{DiracEquation2spinorM}
\end{equation}
Given the above general properties of $\Psi_{n}$, we easily see
that $A^\mu(k)$ with $|k_\perp|\lesssim1/a_B$ and $|k_z|\lesssim
\gamma/a_B$ gives the predominant contribution to the $k$
integration in the right hand side of (\ref{DiracEquation2spinorM}).
As reviewed in the Appendix, in QED the classical field
$A^\mu(vk_z,\vec{k})$ arises from virtual photons with momenta
$\vec{k}$ in the wave-function of the heavier particle in the same
gauge. Therefore, we conclude in the partonic language that those
photons with $|k_\perp|\lesssim 1/a_B$ and $|k_z|\lesssim\gamma/a_B$
are essential for the binding of a bound state moving at velocity
$v$.

This analysis allows us to give a qualitative picture of how the
binding is affected by the presence of the plasma and under which
circumstances the existence of bound states is not allowed. As will
be seen in the next section, the presence of the plasma becomes
manifest through an effective photon mass $\mu$, which causes the
corresponding screening effect. This effective mass provides a
cut-off on the lower limit of the integration on the right hand side
of (\ref{DiracEquation2spinorM}) and, therefore, determines if the
photons responsible for the binding are still available. Following
this argument, the criterion for the dissociation of bound states at
rest in a plasma is $a_{B}\simeq 1/\mu$.

Back to the limit $v\simeq1$, the arguments stated above show that
we can focus only on the contribution to the field from photons
with $k_z\sim\gamma k_\perp$. Before doing this, we need a
simplified Dirac equation in light-cone coordinates which will
enable us to use an uncertainty principle analysis for a fast moving
bound state.

\subsection{A Schr\"{o}dinger-like equation in light-cone gauge in light-cone
coordinates}

In order to determine the existence of fast moving bound states in
the presence of the plasma we would have to solve the Dirac equation
corresponding to that system. For the present case, the Dirac
equation is too complicated to be solved exactly. Nevertheless, we
can get an approximate Schr\"odinger-like equation which will put us
on familiar grounds to make order of magnitude estimates on the
conditions under which bound states are allowed to exist. The major
difficulty comes from the fact that in the rest frame of the charge
the effective field is highly anisotropic in the limit
$v\simeq1$\cite{Chu:1988wh}. In contrast, as showed in Sec.
\ref{effectivelcg}, in the reference frame with the plasma at rest,
the anisotropy of the effective field in light-cone gauge is
predominately due to the Lorentz contraction and can be easily
handled when expressed in light-cone coordinates. This allows us to
neglect the anisotropy of other kinds and greatly simplify the
problem. By using light-cone coordinates and keeping terms up to
$\mathcal{O}(\alpha^{2})$ we are able to derive an equation in
which, by means of the uncertainty principle, we can establish
necessary conditions for the existence of a bound state.

In coordinate space, the Dirac equation
(\ref{DiracEquation2spinorM}) takes the form
\begin{equation}
        \left\{
        \begin{array}{l}
            (p-e_aA)\cdot\sigma\chi=m_a\varphi\\
            (p-e_aA)\cdot\bar{\sigma}\varphi=m_a\chi
        \end{array}
        \right.,\label{DiracEquation2spinorCoor}
\end{equation}
from which we can get two second order differential equations
\begin{equation}
        \left\{
        \begin{array}{l}
            (p-e_aA)\cdot\sigma(p-e_aA)\cdot\bar{\sigma}\varphi=m_a^2\varphi\\
            (p-e_aA)\cdot\bar{\sigma}(p-e_aA)\cdot\sigma\chi=m_a^2\chi
        \end{array}
        \right..
\end{equation}
After some algebra we get
\begin{equation}
       \left\{
        \begin{array}{l}
            \left[(p-e_aA)^2+e_a(\vec{B}+i\vec{E})\cdot\vec{\sigma}\right]\varphi=m_a^2\varphi\\
            \left[(p-e_aA)^2+e_a(\vec{B}-i\vec{E})\cdot\vec{\sigma}\right]\chi=m_a^2\chi
        \end{array}
        \right..\label{DiracEquationFields}
\end{equation}
In the following we only keep terms with expectation value up to
$\mathcal{O}(\alpha^2)$ and neglect terms which are higher order in
$\alpha$. Under this assumption the second term can be ignored in Eq. (\ref{DiracEquationFields}) in
both of the equations given there. We show a detailed calculation of the
potential for the vacuum case in the Appendix which supports this
statement. In the plasma the electric and magnetic fields are
screened and are even weaker than in the vacuum case. Since the two
equations are the same in this approximation, let us focus on one of
them
\begin{equation}
\begin{split}
    (p-e_aA)^2\varphi=m_a^2\varphi.
\end{split}
\end{equation}
Even though the dominant part of the potential is in the transverse
components, the main contribution in the equation above comes from
the $A^{-}$ component since it is enhanced by a $p^{+}$ factor. In
this way we get a Schr\"{o}dinger-like equation in light-cone gauge
in light-cone coordinates
\begin{equation}
p^-\varphi\simeq\left[{p^2_\perp+m_a^2\over
p^+}+e_aA^-\right]\varphi,\label{SchoedingInLC}
\end{equation}
where $p^{\pm}\equiv p^0\pm p^3$. Here we have assumed $p_{\perp}$ is of order $\frac{1}{r}\sim\alpha m_{a}$ and then we have dropped terms which are higher order in $\alpha$. In this paper, we will not solve (\ref{SchoedingInLC})
exactly, but instead we will use the uncertainty principle to
estimate the existence of bound state solutions in the limit
$v\simeq1$. Inspired by (\ref{SchoedingInLC}), we start with
\begin{equation}
\left<p^-\right>\simeq{\left<p^2_\perp\right>+m_a^2\over
\left<p^+\right>}+e_aA^-(\left<r\right>),
\end{equation}
where $\left<p^+\right>=\left[\left<p^{+2}\right>-\Delta
p^{+2}\right]^{1\over
2}\simeq\sqrt{\left<p^{+2}\right>}\left[1-{\left<\Delta
p^+\right>^{2}\over 2\left<p^{+2}\right>}\right]\simeq 2\gamma
m_a\left[1-{\Delta p^{+2}\over 2(2\gamma m_a)^{2}}\right]$, and we have taken
$\sqrt{\left<p^{+2}\right>}\simeq2\gamma m_a$. In the case that
$A^-(r)$ only depends on
$r\equiv\sqrt{x_\perp^2+\gamma^2(x^-)^{2}}$, that is, the system has
a generalized rotational symmetry, we may expect
\begin{equation}
\gamma\left<p^-\right>\simeq{\left<\vec{p}^2\right>\over
2m_a}+{m_a\over 2}+e_a\gamma
A^-(\left<r\right>),\label{SchoedingerInLCUP}
\end{equation}
with $\vec{p}=(p_\perp,\Delta p^+/\gamma)$ and
$\vec{x}=(x_\perp,\gamma x^-)$. Except for the different definitions
of the 3-components of $\vec{x}$ and $\vec{p}$, the physical meaning
of (\ref{SchoedingerInLCUP}) is exactly the same as that used in the
uncertainty principle analysis from the Schr\"{o}dinger equation. In
the vacuum case, we have $A^-={e_b\over 4\pi}{2(1-v)\gamma\over r}$
in light-cone gauge as calculated in the Appendix, and
\begin{equation}
\gamma\left<p^-\right>\simeq{\left<\vec{p}^2\right>\over
2m_a}+{m_a\over 2}-{\alpha\over \left<r\right>},
\end{equation}
in the limit $\gamma\gg1$. By using the uncertainty principle we
have $\gamma\left<p^-\right>\simeq{m_a\over 2}-{\alpha^2m_a\over 2}$
and $p\simeq{1\over r}\simeq\alpha m_a$. This is indeed consistent
with the results obtained from boosted wave functions by keeping
terms up to $\mathcal{O}(\alpha^2)$ in binding energy. Therefore, as
in the limit $v\simeq0$, we can get a Schr\"{o}dinger-like equation
in the limit $v\simeq1$ which allows us to use the uncertainty
principle to estimate the properties of fast moving bound states.

Before calculating the plasma effect on the effective potential, we
give a quantitative estimate about how well the color Coulomb
(perturbative) potential applies to the heavy mesons in the vacuum.
Using the uncertainty principle with the Cornell confining potential
$V(r)=Kr-{\alpha C_F\over r}$\cite{Eichten:1978tg}, we have
\begin{equation}
p={Km_a\over p^2}+\alpha C_Fm_a,
\end{equation}
which tells us that if
\begin{equation}
m_a^2\gg{K\over(\alpha C_F)^3}\simeq{0.2\over(\alpha
C_F)^3}GeV^2,\label{Ispert}
\end{equation}
we may neglect the non-perturbative linear potential responsible
for the confinement in $V(r)$. In this case the discussion about the
electromagnetic bound state given in this section is also valid for
the heavy meson if we replace $\alpha$ with $\alpha C_F$. For the
charm quark, ${0.2\over(\alpha C_F)^3}GeV^2\simeq1.3GeV^2$ with
$m_c^2=1.25^2GeV^2\simeq1.6GeV^2$ and $\alpha(m_c)=0.4$. For the
bottom quark, ${0.2\over(\alpha C_F)^3}GeV^2\simeq10.5GeV^2$ with
$m_b^2=4.7^2GeV^2\simeq22GeV^2$ and $\alpha(m_b)=0.2$. Even though
for the charm and the bottom, equation (\ref{Ispert}) is not
perfectly satisfied, as a parametric estimate in the following
sections we can still take the binding energy
$E_B\simeq\alpha^2C_F^2m_a$.

\subsection{The intrinsic time-scale in a bound state}

From a perturbative point of view, the interaction between the two
quarks in a heavy meson occurs via interchange of gluons. It is
possible to define a typical time within which we can neglect the
interaction between them. This time scale will play an important
role in establishing the appropriate criterion for dissociation due to
multiple scattering.

\begin{figure}
\begin{center}
\includegraphics[width=10cm]{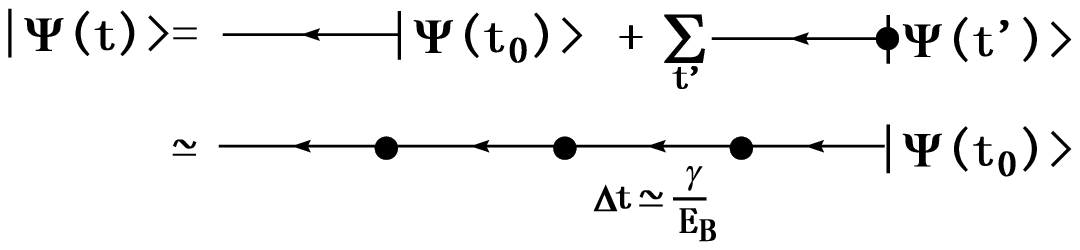}
\end{center}
\caption{The perturbative definition of a heavy meson. Black dots
here represent kicks to quark $a$ from the color field generated by
quark $b$. If $t-t_0\lesssim \gamma/E_B$, the wave packet propagated
by the free propagator represents the predominant contribution to
the wave function at $t$. Approximately, the picture of the
interaction between the two quarks in a heavy meson is that quark
$a$ is kicked once by the color field of quark $b$ within a period
$\Delta t\lesssim\tau_B=\gamma/E_B$ to pull it back into the heavy
meson.}\label{bseFig}
\end{figure}

Let us start with the perturbative definition of the wave
function\cite{Feynman:1949hz},
\begin{equation}
\begin{split}
\Psi_{n\alpha}(x_a)=&\int
d\sigma(x^\prime_a)\left[S^a_F(x_a-x^\prime_a)n\llap{$/$}(x^\prime_a)\right]_{\alpha\rho}
\Psi_{n\rho}(x^\prime_a)\\
&-ie_a\int
d^4x^\prime_a\left[S^a_F(x_a-x^\prime_a)A\llap{$/$}(x^\prime_a)\right]_{\alpha\rho}
\Psi_{n\rho}(x^\prime_a),\label{bse}
\end{split}
\end{equation}
where $d\sigma(x^\prime_a)$ is the volume element of the closed
3-dimensional surface of a region of space time containing $x_a$,
$n^\mu(x^\prime_a)$ is the inward drawn unit normal vector of this
surface at $x^\prime_a$, $A^\mu$ is calculated perturbatively by the
one-gluon exchange approximation and we neglect the linear potential
responsible for the confinement since the masses are assumed much greater
than $\Lambda_{QCD}$. Choosing the integration surface on the first term on the right hand side of (\ref{bse}) as the whole space at two different fixed times, we can relate the wave function at time $t_{a}$ to the values of the wave function at a previous time $t_{0}$. If $\Delta t\equiv t_a-t_0\ll
\tau_B\simeq\gamma{ 1\over \alpha^2C_F^2 m_a}$, the first term on
the right-hand side of (\ref{bse}) gives the predominant
contribution to the wave function at $t_a$. Therefore, by
conservation of probability, one can neglect the second term, that
is, the interaction between these two quarks can be neglected. This
is easy to see in light-cone coordinates, where the free propagator
is \cite{Kogut:1969xa}
\begin{equation}
S_F(x)=\int { d^3p \over (2\pi)^3 } { 1\over 2p^+}\left[
u(p)\bar{u}(p)e^{-ip\cdot x}\Theta(x^+) - v(p)\bar{v}(p) e^{ip\cdot
x} \Theta(-x^+) \right],\label{PropFermLC}
\end{equation}
with $p^-={ p_{\perp}^2+m_a^2\over p^+}$, $\vec{p}=(p_\perp,p^+)$,
and $p^+$ is integrated over the region $0<p^+<\infty$. If typically ${ p_{\perp}^2\over p^+ }\Delta x^+\sim {
\alpha^2C_F^2m_a^2\over \gamma m_a}\Delta t\simeq\Delta t/\tau_B\ll
1$, one can neglect it in the exponentials in (\ref{PropFermLC})
and, therefore, the integration of $\vec{x}^\prime_a$ in the first
term on the right-hand side of (\ref{bse}) simply gives a
$\delta$-function to reproduce the wave function at $t_a$ in the
limit $\gamma\gg1$. Otherwise, if ${ p_{\perp}^2\over p^+ }\Delta
x^+\sim { \alpha^2C_F^2m_a^2\over \gamma m_a}\Delta t\simeq\Delta
t/\tau_B\gg 1$, the first term on the right-hand side of (\ref{bse})
is highly suppressed due to this big term in the exponentials in the
$p_\perp$ integration and, therefore, the second term  on the
right-hand side of (\ref{bse}) contributes predominately by
conservation of probability. Approximately, the picture of
interaction between the two quarks in such a heavy meson is that the
lighter quark interacts with the color field generated by the
heavier quark once every interval $\Delta
t\lesssim\tau_B=\gamma/E_B$. We illustrate this in Fig.
\ref{bseFig}.

\section{Dynamic Debye Screening effect for fast moving bound states}

The next step following our analysis of the Dirac equation would be
to calculate the appropriate potential for a fast moving particle to
plug into equation (\ref{SchoedingerInLCUP}). However, this whole
analysis ignores completely the effect of multiple scattering with
the particles in the medium, which can modify strongly the wave
function of the system. Although in general these two mechanisms
interfere, we will treat them separately in order to determine which
one is dominant and which one should be used to establish a
criterion for existence of bound states in the plasma. This
approximation relies on the fact that screening is a coherent effect
involving a correlated motion of particles in the medium, unlike
multiple scattering which is an incoherent effect caused by random
kicks from uncorrelated scattering centers. In this section, we will
give a detailed calculation for the hot QED plasma, and generalize
the results to the quark-gluon plasma.

\subsection{The photon polarization vector within the HTL approximation}

First, let us calculate the screening effect on the field induced by
the heavy particle, due to the presence of the plasma. This is done
by calculating the retarded photon propagator in thermal field
theory for different regions of momenta. Following the analysis
presented in the previous section, we are mainly interested in those
photons with momenta $k_{z}\sim \gamma k_{\perp}$, which are
essential for the binding of a fast moving bound state. In the
following calculation of the retarded photon propagator, we keep
terms only up to first order in $k_\perp/k_z\sim{1\over\gamma}$.
Moreover, we neglect the modification of $A^\mu$ due to the
appearance of the lighter particle (for a detailed discussion, from
the kinetic theory point of view, about the effective field $A^\mu$
in the rest frame of the particles and the influence of the
appearance of another heavy particle see Ref.\cite{Chu:1988wh}).
Even though the photon polarization vector within the hard thermal
loop (HTL) approximation is well-known (say,
\cite{Kapusta:1989tk,LeBellac:1991cq}), in the following we still
present some details of the calculation, which enables us to see how
well our approximation is in the regions of momenta beyond the HTL
approximation.

Using the notation in Ref. \cite{Kapusta:1989tk}, let us calculate
the photon polarization vector \cite{Kapusta:1989tk,LeBellac:1991cq}
\begin{equation}
\begin{split}
\Pi^{\mu\nu}&=FP_L^{\mu\nu}+GP_T^{\mu\nu}\\
&=e^2T\sum\limits_n\int{d^3q\over (2\pi)^3}{
\mbox{Tr}\left[\gamma^\mu
q\llap{$/$}\gamma^\nu(q\llap{$/$}-k\llap{$/$})\right] \over [
\omega_n^2 +
\omega_{\vec{q}}^2][(\omega_n-\omega)^2+\omega_{\vec{q}-\vec{k}}^2]},\label{PhotonPolarization}
\end{split}
\end{equation}
with $P_L^{\mu\nu}\equiv-g^{\mu\nu}+{k^\mu k^\nu\over
k^2}-P_T^{\mu\nu}$, $P_T^{ij}\equiv\delta_{ij}-\hat{k}_i\hat{k}_j$,
$P_T^{0\mu}=P_T^{\mu 0}=0$ and $\omega_n=(2n+1)\pi T$. From Eq.
(\ref{PhotonPolarization}), we have
\begin{equation}
F={\Pi^{00}\over P_L^{00}}={k^2\over \vec{k}^2}\Pi^{00},\label{DefF}
\end{equation}
and
\begin{equation}
G=-{1\over 2}\left(F+\Pi\right),
\end{equation}
with $\Pi\equiv g_{\mu\nu}\Pi^{\mu\nu}$. After taking the trace and
performing the frequency sums we get,
\begin{equation}\label{Pizerozero}
\begin{split}
\Pi^{00}&=e^{2}\int \frac{d^{3}q}{(2\pi)^{3}} \left\{ \left(1-\frac{E_{1}^{2}-\vec{k}\cdot\vec{q}}{E_{1}E_{2}} \right) \left(1-\tilde{n}(E_{1})-\tilde{n}(E_{2})\right) \left(\frac{1}{k^{0}-E_{1}-E_{2}} - \frac{1}{k^{0}+E_{1}+E_{2}} \right)\right. \\
&\left.+ \left(1+\frac{E_{1}^{2}-\vec{k}\cdot\vec{q}}{E_{1}E_{2}}
\right) \left(\tilde{n}(E_{1})-\tilde{n}(E_{2})\right)
\left(\frac{1}{k^{0}+E_{1}-E_{2}} - \frac{1}{k^{0}-E_{1}+E_{2}}
\right)\right\},
\end{split}
\end{equation}
and
\begin{equation}\label{Pi}
\begin{split}
\Pi&=-2e^{2}\int \frac{d^{3}q}{(2\pi)^{3}} \left\{ \left(1+\frac{E_{1}^{2}-\vec{k}\cdot\vec{q}}{E_{1}E_{2}} \right) \left(1-\tilde{n}(E_{1})-\tilde{n}(E_{2})\right) \left(\frac{1}{k^{0}-E_{1}-E_{2}} - \frac{1}{k^{0}+E_{1}+E_{2}} \right)\right. \\
&\left.+ \left(1-\frac{E_{1}^{2}-\vec{k}\cdot\vec{q}}{E_{1}E_{2}}
\right) \left(\tilde{n}(E_{1})-\tilde{n}(E_{2})\right)
\left(\frac{1}{k^{0}+E_{1}-E_{2}} - \frac{1}{k^{0}-E_{1}+E_{2}}
\right)\right\},
\end{split}
\end{equation}
where $\tilde{n}(E)=\frac{1}{e^{\beta E}+1}$, $E_{1}=|\vec{q}|$, and
$E_{2}=|\vec{q}-\vec{k}|$.

In the HTL approximation, that is, $k_z, k_\perp\ll T$, we get the
well-known results
\begin{equation}
\begin{split}
\Pi^{00}&\simeq 2e^{2}\int \frac{d^{3}q}{(2\pi)^{3}}
{d\tilde{n}(q)\over dq}\left(2-{k^0\over k^0+k\cos\theta}-{k^0\over
k^0-k\cos\theta}\right)\\
&=-\mu^2\left(2-{k^0\over k}\ln{{{k^0\over k}+1\over {k^0\over
k}-1}}\right),
\end{split}
\end{equation}
and
\begin{equation}
\begin{split}
\Pi&\simeq4e^{2}\int \frac{d^{3}q}{(2\pi)^{3}} {1\over
q}\left(1-2\tilde{n}(q)\right)\\
&\rightarrow-4e^{2}\int \frac{d^{3}q}{(2\pi)^{3}} {1\over
q}2\tilde{n}(q)=-2\mu^2,
\end{split}
\end{equation}
where we have dropped the $T$-independent divergent term and we have
taken $\mu^2\equiv{e^2T^2\over 6}$. When making use of the classical
current induced by a fast moving particle, we introduce a delta
function of the form $\delta(k^{0}-vk_{z})$, so we can safely
replace $k^0=vk_z$ in the calculation of the propagator. In our case
$k_z\simeq\gamma k_\perp$, by keeping only the first order of
$k_\perp\over k_z$, we have
\begin{equation}
F\simeq-\mu^2{k^2\over\vec{k}^2}\left(2-{k^0\over k}\ln{{{k^0\over
k}+1\over {k^0\over
k}-1}}\right)\simeq2\mu^2\gamma^{-2}\ln{\gamma}\xrightarrow{v\to
1}0,\label{PiHTL}
\end{equation}
and, therefore,
\begin{equation}
G\simeq\mu^2.\label{GHTL}
\end{equation}

\subsection{The photon polarization vector beyond the HTL approximation}

As explained in previous sections, photons with $k_{z}\simeq
\gamma/a_{B}$ play an important role in the calculation of bound
state wave functions, and since we are interested in fast moving
particles, $k_{z}$ might be comparable with or even larger than $T$
in the limit $v\simeq1$. On the other hand, the transverse
components of the photon momenta are not necessarily large. The
relevant region for the transverse momenta is $\sim 1/a_{B}$ and in
particular we would like to focus on the region $1/a_{B}\sim \alpha^{1/2}T$,
which means we still have the condition $k_{\perp}\ll T$. In the
following, we give a detailed calculation of the photon polarization
vector in the limit $k_\perp\ll T$ and $k_z\gg T$.

Let us calculate $F$ and $G$ for this region of momenta. Since we
are considering the case $|\vec{k}|\gg T$, the predominant
contributions to the integrals in (\ref{Pizerozero}) and (\ref{Pi})
come from the separate regions with $|\vec{q}|\sim T$ or
$|\vec{q}-\vec{k}|\sim T$. Both regions give the same contribution
since the integrands are unchanged under
$\vec{q}\to\vec{k}-\vec{q}$. Taking into account the contribution
from both regions we get
\begin{equation}
\begin{split}
F={k^2\over \vec{k}^2}4e^2\int
\frac{d^{3}q}{(2\pi)^{3}}{\tilde{n}(q)\over q}{1+\cos^2\theta\over
1-\cos^2\theta}={k^2\over \vec{k}^2}\mu^2\int_0^1
d\cos\theta{1+\cos^2\theta\over 1-\cos^2\theta}.
\end{split}
\end{equation}
The angular integration has a collinear divergence that is cutoff by
the $k_{\perp}$ term neglected in the approximation above and by the
mass of the constituents of the plasma (neglected in our calculation
for high $T$). Therefore, as in the case $k_z, k_\perp\ll T$ in
(\ref{PiHTL}), this
 logarithmic divergence must be proportional to $\ln\gamma$, that is,
\begin{equation}
F\sim\gamma^{-2}\ln{\gamma}.
\end{equation}
Similarly, the predominant contributions of $q$-integration from the
(separate) regions with $|\vec{q}|\sim T$ or $|\vec{q}-\vec{k}|\sim
T$ gives us
\begin{equation}
\Pi=-2\mu^2.
\end{equation}
Accordingly,
\begin{equation}
\begin{split}
G=\mu^2\left(1-{1\over 2}{k^2\over \vec{k}^2}\int_0^1
d\cos\theta{1+\cos^2\theta\over 1-\cos^2\theta}\right) &\simeq\mu^2.
\end{split}
\end{equation}
Therefore, in the case $k_\perp\ll T$, we have $F\simeq0$ and
$G\simeq\mu^2$ in the limit $k_z\gg T$ as well as in the limit
$k_z\ll T$. This justisfies $F\simeq0$ and $G\simeq\mu^2$ in the
case $k_\perp\ll T$ in the following calculation even though the
region with $k_z\sim T$ is difficult to evaluate analytically.

\subsection{The effective field in light-cone gauge}\label{effectivelcg}

To use (\ref{SchoedingerInLCUP}) for a fast moving bound state in
the plasma, we need to evaluate the retarded photon propagator in
light-cone gauge. By only keeping terms up to first order in
$k_\perp/k_z\simeq\gamma^{-1}$, we have
\begin{equation}
\begin{split}
&\left[D_{LC}D_{LC}\right]^{\mu\nu}\equiv D_{LC\rho}^\mu
D_{LC}^{\rho\nu}\simeq{-i\over k^2}D_{LC}^{\mu\nu},\label{ProdofDLC}
\end{split}
\end{equation}
and
\begin{equation}
\begin{split}
&\left[-iD_{LC}\Pi D_{LC}\right]^{\mu\nu}\equiv -iD^{\mu\rho}_{LC}\Pi_{\rho\sigma}D^{\nu\sigma}_{LC}\\
&\simeq{i\mu^2\over
\left[k^2\right]^2}\left[P_T^{\mu\nu}-{1\over\eta\cdot
k}\left(k^\mu\eta_\sigma P_T^{\nu\sigma}+k^\nu\eta_\sigma
P_T^{\mu\sigma}\right)+{k^\mu k^\nu\over(\eta\cdot
k)^2}\eta^{\rho}\eta^{\sigma}P^T_{\rho\sigma}\right]\simeq{\mu^2\over
k^2} D^{\mu\nu}_{LC},\label{SeriesofDR}
\end{split}
\end{equation}
where $D_{LC}^{\mu\nu}$, defined in (\ref{DRInLC}), is the vacuum light-cone gauge propagator. Using
(\ref{ProdofDLC}) and (\ref{SeriesofDR}), we get the full photon
propagator
\begin{equation}
\begin{split}
D^{\mu\nu}_R&= D^{\mu\nu}_{LC}+\left[-iD_{LC}\Pi
D_{LC}\right]^{\mu\nu} +\left[(-i)^2D_{LC}\Pi D_{LC}\Pi
D_{LC}\right]^{\mu\nu}
+\cdot\cdot\cdot\\
&\simeq-{i\over k^2-\mu^2}\left[g^{\mu\nu}-{\eta^\mu k^\nu+\eta^\nu
k^\mu\over\eta\cdot k}\right].\label{DRInQEDInLC}
\end{split}
\end{equation}

The corresponding effective field is given by Maxwell's equations
\begin{equation}
A^\mu(k)=-iD_R^{\mu\nu}(k) j_\nu\simeq 2\pi e_b\delta(\omega-vk_z)(
{v-1 \over k^2-\mu^2},{\vec{k}_\perp\over k_z( k^2-\mu^2)},{1-v\over
k^2-\mu^2}),\label{AeffectiveinC}
\end{equation}
and in coordinate space we have
\begin{equation}
A^-(x)={e_b\over 4\pi}2(1-v)\gamma{e^{-\mu r}\over
r},\label{AinLCPlas}
\end{equation}
and
\begin{equation}
\begin{split}
\vec{A}_\perp(x)&={i\over2}e_b\bigtriangledown_\perp\int
{d^3k\over(2\pi)^3}e^{-ik\cdot
x}{1\over\left[k_\perp^2+\gamma^{-2}k_z^2+\mu^2\right]}
\left({1\over k_z+i\epsilon}+{1\over k_z-i\epsilon}\right)\\
&\simeq
i{e_b\over4\pi}\left[\Theta(x^-)-\Theta(-x^-)\right]\bigtriangledown_\perp\int
dk_\perp
k_\perp{J_0(k_\perp x_\perp)\over k_\perp^2+\mu^2}\\
&\simeq{e_b\over 4\pi}{ \vec{x}_\perp\over x_\perp^2}\mu x_\perp
K_1(\mu x_\perp)\left[\Theta(x^-)-\Theta(-x^-)\right],
\end{split}
\end{equation}
where
$r\equiv\sqrt{x^2_\perp+\gamma^2(z-vt)^2}\simeq\sqrt{x^2_\perp+\gamma^2(x^{-})^2}$
and in the calculation of $A_\perp(x)$ we have only picked up the
poles at $k_z=\pm i\epsilon$.

In kinetic theory, the effective field $\tilde{A}^{\mu}(k)$ is found
to be strongly anisotropic for $v\simeq1$ in the rest frame of the
charge \cite{Chu:1988wh}, which seems to contradict the results in
light-cone gauge. However, we shall see that after some gauge
transformation and going back to the rest frame of the plasma, the
anisotropy is suppressed as inverse powers of $\gamma$ in contrast
with that due to the Lorentz contraction. In the limit $v\simeq1$,
the effective field $\tilde{A}^{\mu}(k)$ in the rest frame of the
charge calculated in Ref. \cite{Chu:1988wh} is
\begin{equation}
\begin{split}
\tilde{A}^0(k)&\simeq2\pi e_{b}\delta(\omega)\left[{1\over
\vec{k}^2+\mu^2}-{1\over\cos^2\theta}\left({1\over
\vec{k}^2+\mu^2}-{1\over \vec{k}^2}\right)\right],\\
\tilde{A}^1(k)&\simeq-2\pi
e_{b}\delta(\omega)\tan\theta\cos\phi\left({1\over
\vec{k}^2+\mu^2}-{1\over \vec{k}^2}\right),\\
\tilde{A}^2(k)&\simeq-2\pi
e_{b}\delta(\omega)\tan\theta\sin\phi\left({1\over
\vec{k}^2+\mu^2}-{1\over \vec{k}^2}\right),\\
\tilde{A}^3(k)&\simeq2\pi e_{b}\delta(\omega)\tan^2\theta\left({1\over
\vec{k}^2+\mu^2}-{1\over \vec{k}^2}\right),\label{AefffromKin}
\end{split}
\end{equation}
with
$k^\mu=(\omega,|\vec{k}|\cos\phi\sin\theta,|\vec{k}|\sin\phi\sin\theta,|\vec{k}|\cos\theta)$.
Equ. (\ref{AefffromKin}) is the same as calculated in the above
approximation $F\simeq0$ and $G\simeq\mu^2$ in covariant gauge.
After the gauge transformation
$\tilde{A}^\mu(k)\rightarrow\tilde{A}^\mu(k)+k^\mu\Lambda(k)$ with
\begin{equation}
\Lambda(k)=2\pi e_{b}\delta(\omega){1\over
|\vec{k}|\cos\theta}\left({1\over \vec{k}^2+\mu^2}-{1\over
\vec{k^2}}\right),
\end{equation}
we have
\begin{equation}
\begin{split}
\tilde{A}^0(k)&\simeq2\pi e_{b}\delta(\omega)\left[{1\over
\vec{k}^2+\mu^2}-{1\over\cos^2\theta}\left({1\over
\vec{k}^2+\mu^2}-{1\over \vec{k}^2}\right)\right],\\
\tilde{A}_\perp(k)&\simeq0,\\
\tilde{A}^3(k)&\simeq2\pi e_{b}\delta(\omega){1\over
\cos^2\theta}\left({1\over \vec{k}^2+\mu^2}-{1\over
\vec{k}^2}\right).
\end{split}
\end{equation}
Back to the rest frame of the plasma, it gives
\begin{equation}
\begin{split}
A^0(k)&\simeq-2\pi e_{b}\delta(\omega-vk_z){1\over k^2-\mu^2},\\
A_\perp(k)&\simeq0,\\
A^3(k)&\simeq-2\pi e_{b}\delta(\omega-vk_z){1\over k^2-\mu^2},
\end{split}
\end{equation}
showing that the anisotropy is suppressed as inverse powers of
$\gamma$ in the rest frame of the plasma. We have shown that the
effective field of a fast moving charge has a smooth limit as
$v\rightarrow 1$. Moreover, by using light-cone gauge in the rest
frame of the plasma, one can obtain an effective field in which the
anisotropy is predominantly due to Lorentz contraction. This allows
us to use (\ref{SchoedingerInLCUP}) to obtain a criterion for the
dissociation of fast moving heavy mesons due to dynamic Debye
screening.

\subsection{Dissociation of heavy mesons due to dynamic Debye screening}

In the HTL approximation, one can obtain the gluon self-energy
simply by taking $\mu^2={1\over 6}g^2T^2(N_c+{N_f\over2})$  in
(\ref{PiHTL}) and (\ref{GHTL})\cite{LeBellac:1991cq}, where $N_f$ is
the number of massless flavors of quarks in the quark-gluon plasma.
Since this result does not depend on the gauge-fixing
\cite{Kobes:1990dc}, by replacing $e_b$ with $-g$, we get the
effective gluon field in light-cone gauge
\begin{equation}
A^-(x)=-{g\over 4\pi}2(1-v)\gamma{e^{-\mu r}\over
r},\label{AinLCPlasQCD}
\end{equation}
and
\begin{equation}
A_\perp(x)\simeq-{g\over 4\pi}{ \vec{x}_\perp\over x_\perp^2}\mu
x_\perp K_1(\mu x_\perp)\left[\Theta(x^-)-\Theta(-x^-)\right].
\end{equation}
Now, we are ready to discuss the criterion for the dissociation of
fast moving heavy mesons in the quark-gluon plasma by using the
uncertainty principle in light-cone gauge. Inserting
(\ref{AinLCPlas}) into (\ref{SchoedingerInLCUP}), we have
\begin{equation}
\gamma \left<p^-\right>\simeq{\vec{p}^2\over 2m_a}+{m_a\over
2}-2(1-v)\gamma^2{\alpha C_Fe^{-\mu r}\over r}\simeq{\vec{p}^2\over
2m_a}+{m_a\over 2}-{\alpha C_Fe^{-\mu r}\over
r},\label{SchoedingerInLCUPInPlas}
\end{equation}
in the limit $v\simeq 1$. Assuming $r=1/p$ and minimizing
(\ref{SchoedingerInLCUPInPlas}), we have
\begin{equation}
{\mu\over \alpha C_Fm_a}=x(1+x)e^{-x},\label{EquaOfUP}
\end{equation}
with $x\equiv\mu/p$. Eq. (\ref{EquaOfUP}) has solutions only if $\mu
a_B\equiv{\mu\over \alpha C_Fm_a}\leq0.84$ (see
\cite{Matsui:1986dk}). Therefore, we can use the uncertainty
principle argument for the dissociation of fast moving heavy mesons
in the plasma in the same way as it was used for bound states at
rest \cite{Matsui:1986dk}. The corresponding criterion for the
dissociation of fast moving heavy mesons, based on a screening
analysis, is
\begin{equation}
a_B\simeq 1/\mu.\label{DDS}
\end{equation}
Note, the anisotropy in the effective field seen in the rest frame
of the charge could imply a more efficient screening than that in
the case with the charge almost at rest with the plasma, but the
modification of this criterion on the right hand of (\ref{DDS})
should not depend on inverse powers of $\gamma$.

\section{Dissociation of Heavy Mesons due to Multiple scattering}

In this section, we give a parametric estimate of the criterion for
the dissociation of heavy mesons due to multiple scattering in terms
of the saturation momentum $Q_s$, which is a characteristic property
of any QCD media\cite{Mueller:1999yb,Mueller:1999wm}.
Quantitatively, the physical meaning of $Q_s$ is that the gluon
distribution of the target is dense as seen by probes with
virtuality $q_\perp\ll Q_s$, but it is dilute as seen by probes with
high virtuality. When a heavy meson with a size $a_B$ travels in the
quark-gluon plasma, $1/a_B$ naturally plays the role of the virtuality
$q_\perp$. If $a_B\gg 1/Q_s$, the meson will break up, that is,
the medium looks opaque to the meson. This picture is confirmed by
detailed calculations in Ref. \cite{MultipleScattering}.

This picture can also be justified by the argument of the transverse
momentum broadening of the two quarks in a heavy meson. As showed in
Fig. \ref{bseFig}, within a period $\tau_B$, the two particles in a
heavy meson propagate in the medium like two free quarks and pick up
the transverse momentum broadening $\left<\Delta p_\perp^2\right>$,
which is equal to the saturation momentum squared
$Q_s^2$\cite{Baier:1996kr}. If $\left<\Delta
p_\perp^2\right>=Q_s^2\gg1/a_B^2$, that is, the two quarks pick up
transverse momenta greater than the typical momentum in a bound
state, the meson will break up. Therefore, the criterion for the
dissociation of a heavy meson due to multiple scattering in a medium
is parametrically $a_B\simeq 1/Q_s$. This picture applies to cold
matter as well as to hot matter\cite{MultipleScattering}. In the
following, we shall give a parametric analysis of the criterion for
the dissociation of heavy mesons in a hot quark-gluon plasma in the
case that the successive scatterings are uncorrelated.

In a hot quark-gluon plasma modeled by uncorrelated scattering
centers, for a fast moving quark, $Q_s^2$ has the following simple
form\cite{Baier:1996kr}
\begin{equation}
Q_s^2\simeq {L\over \lambda}\mu^2\ln{T^2\over
\mu^2}=L\rho\sigma\mu^2\ln{T^2\over \mu^2}\simeq\alpha^2
C_F(N_c+{N_f\over2})T^3L\ln{1\over\alpha_{eff}},\label{QsforRelQua}
\end{equation}
where we have taken the number density $\rho\simeq
(N_c+{N_f\over2})T^3$, $\sigma={4\pi\alpha^2\over\mu^2}C_F$ and $\mu^2={1\over
6}(N_c+{N_f\over2})g^2T^2\simeq\alpha(N_c+{N_f\over2})T^2\equiv\alpha_{eff}T^2$.
In a finite plasma with a length $L\lesssim\tau_B$, the criterion
for the dissociation of heavy mesons is $\alpha^2
C_F(N_c+{N_f\over2})\ln{1\over\alpha_{eff}}T^3L\simeq{1\over
a_B^2}$. In an infinite quark-gluon plasma, the time scale
$\tau_B={\gamma\over E_B}\simeq{\gamma a_B\over \alpha C_F}$ plays
the role of the length $L$ in the definition of $Q_s$, and our
criterion for dissociation of heavy mesons $Q_s^2\simeq 1/a_B^2$
gives $a_B\simeq {1\over [\gamma\alpha_{eff}\ln{1\over
\alpha_{eff}}]^{1\over 3}T}$ for the quark-gluon plasma, in
contrast with that due to the screening effect
$a_B\simeq{1\over\mu}\simeq{1\over\alpha_{eff}^{1\over2}T}$.

If the successive scatterings between the mesons and the plasma
constituents are essentially independent of each other, the
criterion $a_B\simeq {1\over [\gamma\alpha_{eff}\ln{1\over
\alpha_{eff}}]^{1\over 3}T}$ is also true for the dissociation
of heavy mesons almost at rest with the plasma. Quantitatively, in a
hot plasma this means the collision time
$\tau_c\simeq\lambda={1\over\rho\sigma}\simeq{1\over\alpha C_F T}\gg
1/\mu$, which is equivalent to $1\sim{N_c+{N_f\over2}\over
C_F^2}\gg\alpha$. In this case, non-relativistic quarks pick momentum broadening
symmetrically in each direction due to the uncorrelated random
kicks from plasma constituents in the same way as the transverse
momentum broadening of relativistic quarks. Therefore, we would
expect that the momentum broadening of non-relativistic quarks
should have a similar form as (\ref{QsforRelQua}).

Since uncorrelated multiple scattering implies $\alpha\ll1$, we can
use the leading-log approximation\cite{Arnold:2000dr} in our
discussion about the criterion for the dissociation of
non-relativistic heavy mesons. The mean-squared momentum transfer
per unit time between a non-relativistic heavy quark and the plasma
is calculated in Ref. \cite{Moore:2004tg} and by keeping only the
leading log terms, we have
\begin{equation}
{d\over dt}\left<(\Delta p)^2\right>\simeq {4\pi\over 3}{\alpha^2
C_F}(N_c+{N_f\over2})\ln{T^2\over \mu^2}T^3.\label{Deltapt^2}
\end{equation}
Since Eq. (\ref{Deltapt^2}) is time-independent,
\begin{equation}
\left<(\Delta p)^2\right>\simeq {4\pi\over 3}{\alpha^2
C_F}(N_c+{N_f\over2})\ln{T^2\over \mu^2}T^3\Delta t.\label{Deltap^2}
\end{equation}
Taking $\Delta t\simeq\tau_B\simeq{1\over E_B}$ and $\left<(\Delta
p)^2\right>\simeq 1/a_B^2$, we get $a_B\simeq {1\over
[\alpha_{eff}\ln{1\over \alpha_{eff}}]^{1\over 3}T}$. Therefore, we conclude that $a_B\simeq
{1\over [\gamma\alpha_{eff}\ln{1\over \alpha_{eff}}]^{1\over 3}T}$
is a parametric criterion for the dissociation of a heavy meson due
to uncorrelated multiple scattering in an infinite plasma. This criterion
was also obtained in Ref.\cite{Laine:2008cf} in the case of heavy
quarkonia at rest with the plasma by effective field theory
techniques, which was first obtained (without the logarithm) in
\cite{Escobedo:2008sy}. At high
energies, heavy mesons can dissociate even at much lower temperature
$T$ before dynamic Debye screening effect plays an important role.

\section*{ACKNOWLEGEMENT}
We are indebted to Prof. A. H. Mueller for helpful and illuminating
discussions. B. W. is especially grateful to Prof. Bo-Qiang Ma for
helpful suggestions and other support. We would also like to
acknowledge Cyrille Marquet and Bo-Wen Xiao for helpful discussions.
We thank M. Laine for drawing our attention to
\cite{Laine:2008cf,Escobedo:2008sy}.

\section*{APPENDIX: The Classical Field $A_{cl}
^\mu$ in Coulomb and Light-cone Gauges}

In this appendix, we compare the field $A^\mu$ for a charge $e_b$
moving with a velocity $v$ along the $z$ direction in classical
electrodynamics and the classical field $A_{cl}^\mu$ for a charged
particle  with $p^\mu=p^0(1,0_\perp,v)$ and charge $e_b$ in the
partonic picture in QED. In classical electrodynamics, the field is
calculated using Maxwell's equations with a classical current. In
momentum space, we have
\begin{equation}
A^\mu(k)=-iD^{\mu\nu}(k)j_\nu(k),
\end{equation}
and
\begin{equation}
j^\mu(k)=2\pi e_b\delta(\omega-vk_z)\left(1,0_\perp,v\right).
\end{equation}

(i) Coulomb gauge $\bigtriangledown\cdot\vec{A}=0$

The photon propagator in Coulomb gauge is
\begin{equation}
D_C^{\mu\nu}(k)=
    \left(\begin{array}{cc}
        { i\over \vec{k}^2}&0\\
        0&{ iP_T^{ij}\over k^2}
    \end{array}\right)
\end{equation}
with $P_T^{ij}\equiv\delta_{ij}-\hat{k}_i\hat{k}_j$. The classical electromagnetic field is
\begin{equation}
A^\mu(k )=-iD_C^{\mu\nu}(k) j_\nu(k)=2\pi e_b\delta(\omega-vk_z)( { 1
\over \vec{k}^2},{-vP_T^{i3}\over k^2}).\label{AinC}
\end{equation}
In the limit $k_z\gg k_\perp$ and $v\simeq1$, we have
\begin{equation}
P_T^{i_\perp 3}\simeq - {k_\perp^{i_\perp}\over k_z}, P_T^{33}\simeq
{k_\perp^2\over k_z^2},
\end{equation}
and
\begin{equation}
A^\mu(k)\simeq 2\pi e_b\delta(\omega-vk_z)( { 1 \over
k_z^2},-{\vec{k}_\perp\over k_z k_\perp^2},{1\over k_z^2}),
\end{equation}
which has the same transverse components as (\ref{AinLCv1}).

(ii) Light-cone gauge $A^+=0$

The photon propagator is
\begin{equation}
D_{LC}^{\mu\nu}(k)=-{i\over k^2}\left[g^{\mu\nu}-{\eta^\mu
k^\nu+\eta^\nu k^\mu\over\eta\cdot k}\right],\label{DRInLC}
\end{equation}
and
\begin{equation}
A^\mu(k)=-iD_{LC}^{\mu\nu}(k)j_\nu(k)=2\pi e_b\delta(\omega-vk_z)(- {1-v
\over k^2},{\vec{k}_\perp\over k_z k^2},{1-v \over
k^2}),\label{AinLC}
\end{equation}
where $\eta^\mu={1\over\sqrt{2}}(1,0,0,-1)$. In the limit
$v\simeq1$, we have
\begin{equation}
A^\mu(k)\rightarrow 2\pi
e_b\delta(\omega-vk_z)(0,-{\vec{k}_\perp\over k_z
k_\perp^2},0),\label{AinLCv1}
\end{equation}
which is the same as the Weizs\"{a}cher-Williams field calculated in
the light-cone wave function of particle $b$ in the partonic picture
\cite{Mueller:1999yb}.

In QED, the field is quantized and virtual photons appear in the
wave function of a charged particle to give rise to the classical
field. In this appendix, we use the notations in Ref.
\cite{Peskin:1995ev} and the classical field is defined
\begin{figure}
\begin{center}
\includegraphics[width=8cm]{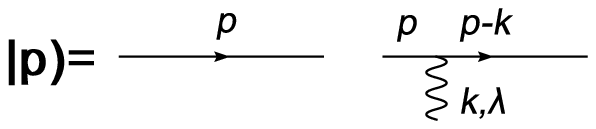}
\end{center}
\caption{The dressed wave function of a charged
particle}\label{DresWaveFuncFig}
\end{figure}
by\cite{Mueller:1999yb}
\begin{equation}
A^\mu_{cl}(\vec{x})=\int{d^3p^\prime\over
2E_{\vec{p}^\prime}(2\pi)^3}\left(p^\prime\right|\hat{A}^\mu(\vec{x})\left|p\right),\label{ClasFiel}
\end{equation}
where $\hat{A}^\mu(\vec{x})$ is the quantized photon field and $|p)$
is the dressed wave function of the charge particle as showed in
Fig. \ref{DresWaveFuncFig}
\begin{equation}
|p)=|p\rangle+\sum\limits_{\lambda=\pm}\int{d^3k\over(2\pi)^3}\Psi_\lambda(k)\left|p-k;k,\lambda\right>,\label{DresWaveFunc}
\end{equation}
with
\begin{equation}
\begin{split}
&2k^0(2\pi)^32E_{\vec{p}-\vec{k}}\Psi_\lambda(k)\delta(\vec{p}-\vec{p}^{\prime})={\left<p^\prime-k;k,\lambda\right|H_I\left|p\right>\over
p^0-(p-k)^0-k^0}\\
&=e_b(2\pi)^3\delta(\vec{p}-\vec{p}^{\prime}){\bar{u}(p-k)\epsilon\llap{$/$}^{*}_\lambda(k)
u(p)\over p^0-(p-k)^0-k^0}.
\end{split}
\end{equation}
Inserting (\ref{DresWaveFunc}) into (\ref{ClasFiel}), we have
\begin{equation}
A_{cl}^\mu(\vec{x})=\sum\limits_{\lambda=\pm}\int
{d^3k\over(2\pi)^3}e^{i\vec{k}\cdot
\vec{x}}\epsilon^\mu_\lambda(k)\Psi_\lambda(k)\equiv\int
{d^3k\over(2\pi)^3}e^{i\vec{k}\cdot \vec{x}}A_{cl}^\mu(k).
\end{equation}
Here, we only keep the positive-energy part of $A_{cl}^\mu$.

The classical field $A_{cl}^\mu$ in light-cone gauge is
well-known\cite{Mueller:1999yb}. And we shall calculate it in
Coulomb gauge as another example to illustrate the correspondence
between $A^\mu$ and $A^{\mu}_{cl}$. In Coulomb gauge,
\begin{equation}
\sum\limits_{\lambda=\pm}\epsilon^\mu_\lambda\epsilon^{\nu*}_\lambda=P^{\mu\nu}_T.
\end{equation}
Assuming $p^0\simeq p_z\gg k^0$, we have
\begin{equation}
\Psi_\lambda(k)\simeq-{e_b\over
2}{\epsilon^0_\lambda-v\epsilon^3_\lambda\over
\vec{k}^2(1-v\hat{k}_z)}\simeq
-e_b{\epsilon^0_\lambda-v\epsilon^3_\lambda\over
\vec{k}^2(1-v^2\hat{k}^2_z)}=e_b{\epsilon^0_\lambda-v\epsilon^3_\lambda\over
k^2},
\end{equation}
and
\begin{equation}
A^{i}_{cl}(\vec{k})\simeq e_b{P_T^{i0}-vP_T^{i3}\over k^2}\simeq
e_b{-vP_T^{i3}\over k^2},\label{AinQFT}
\end{equation}
with $k^2=v^2k_z^2-\vec{k}^2$, which, if multiplied by
$2\pi\delta(\omega-vk_z)$, is the same as the vector potential in
(\ref{AinC}). Therefore, both calculations in Coulomb and light-cone
gauges illustrate the fact that the field $A^\mu(\omega,\vec{k})$
with $\omega=vk_z$ in a certain gauge in classical electrodynamics
arises from the virtual photon with momenta
$k^\mu=(|\vec{k}|,\vec{k})$ in the same gauge in QED.

Since the Dirac equation is more convenient to solve in coordinate
space, we shall evaluate the field $A^\mu(x)$ by the Fourier
transformation of (\ref{AinC}) and (\ref{AinLC}). In Coulomb gauge,
$A^\mu(x)$ is
\begin{equation}
\begin{split}
A^\mu(x)&=\int{d^4k\over (2\pi)^4}e^{-ik\cdot x}A^\mu(k)\\
&=e_b\left({1\over 4\pi r},v{\partial^2\over \partial
\vec{x}_\perp\partial z}\Lambda(x),v\left({\gamma\over 4\pi
\tilde{r}}+{\partial^2\over
\partial z^2}\Lambda(x)\right)\right),\label{AinCoordinateDef}
\end{split}
\end{equation}
where $r\equiv\sqrt{x_\perp^2+(z-vt)^2}$,
$\tilde{r}\equiv\sqrt{x_\perp^2+\gamma^2(z-vt)^2}$ \footnote{Note,
in Sec. II and Sec. III, this is defined as $r$.}, and
\begin{equation}
\begin{split}
\Lambda(x)&\equiv\int{d^3k\over (2\pi)^3}{e^{-ik\cdot x}\over
\vec{k}^2(k_\perp^2+\gamma^{-2}k_z^2)}=\int{d^3k\over
(2\pi)^3}\int_0^1 d\eta{e^{-ik\cdot x}\over
[k_\perp^2+(\eta+(1-\eta)\gamma^{-2})k_z^2]^2}\\
&=\int_0^1 d\eta\int{dk_\perp dk_z k_\perp\over
(2\pi)^2}{e^{ik_z(z-vt)} J_0(k_\perp x_\perp)\over
[k_\perp^2+(\eta+(1-\eta)\gamma^{-2})k_z^2]^2}\\
&=\int_0^1 d\eta\int{dk_z \over 8\pi^2}e^{ik_z(z-vt)}{x_\perp
K_1(\sqrt{(\eta+(1-\eta)\gamma^{-2})k_z^2 x_\perp^2})\over
\sqrt{(\eta+(1-\eta)\gamma^{-2})k_z^2}}\\
&=-\int_0^1 { d\eta \over 8\pi
}{\sqrt{(\eta+(1-\eta)\gamma^{-2})x_\perp^2 +(z-vt)^2}\over
{\eta+(1-\eta)\gamma^{-2}}}\\
&={1\over
4\pi}{r-\gamma^{-1}\tilde{r}-(z-vt)\mbox{ArcTanh}\left({r\over
z-vt}\right)+(z-vt)\mbox{ArcTanh}\left({\gamma^{-1}\tilde{r}\over
z-vt}\right)\over \gamma^{-2}-1}.\label{Lambda}
\end{split}
\end{equation}
Inserting (\ref{Lambda}) into (\ref{AinCoordinateDef}), we obtain
\begin{equation}
A^\mu(x)={e_b\over 4\pi}\left( { 1 \over r},{1\over v}{
\vec{x}_\perp\over x_\perp^2}(z-vt)\left(\gamma{1\over \tilde{r}}-{
1\over r}\right),{ 1\over v}\left({ 1\over r}-{ 1\over \gamma}{
1\over\tilde{r}}\right)\right).\label{AinCoordinate}
\end{equation}

In light-cone gauge, in coordinate space,
\begin{equation}
\begin{split}
\vec{A}_\perp(x)&=-e_b\int
{d^3k\over(2\pi)^3}e^{-ik\cdot x}{\vec{k}_\perp\over\left[k_\perp^2+\gamma^{-2}k_z^2\right]k_z}\\
&=ie_b\bigtriangledown_\perp\int
{d^3k\over(2\pi)^3}e^{-ik\cdot x}{1\over\left[k_\perp^2+\gamma^{-2}k_z^2\right]k_z}\\
&=ie_b\bigtriangledown_\perp\int
{dk_zdk_\perp k_\perp\over(2\pi)^2}e^{ik_z(z-vt)}{J_0(k_\perp x_\perp)\over\left[k_\perp^2+\gamma^{-2}k_z^2\right]k_z}\\
&=ie_b\bigtriangledown_\perp\int
{dk_z\over(2\pi)^2}{e^{ik_z(z-vt)}\over
k_z}K_0(\sqrt{k_z^2\gamma^{-2} x_\perp^2})\\
&=-{e_b\over 4\pi}\bigtriangledown_\perp \mbox{ArcSinh}{\gamma (z-vt)\over x_\perp}\\
&={e_b\over 4\pi}{\vec{x}_\perp\over x^2_\perp}{\gamma (z-vt)\over
\tilde{r}},
\end{split}
\end{equation}
and
\begin{equation}
A^\mu(x)={e_b\over
4\pi}\left((1-v)\gamma{1\over\tilde{r}},{\vec{x}_\perp\over
x^2_\perp}{\gamma (z-vt)\over
\tilde{r}},-(1-v)\gamma{1\over\tilde{r}}\right).
\end{equation}
In the limit $v\rightarrow1$,
\begin{equation}
A^\mu(x)={e_b\over 4\pi}\left(0,{\vec{x}_\perp\over
x^2_\perp}\left[\Theta(x^-)-\Theta(-x^-)\right],0\right),
\end{equation}
which is the same as $A_{cl}^\mu(x)$ in Ref.\cite{Mueller:1999yb}.

It is interesting to notice the big difference of the role played by
different components of the classical field $A^\mu(x)$ in light-cone
gauge in scattering processes at high energies and in fast moving
bound systems. In a scattering process at high energies, if we take
the right-mover as a classical current, since the other particle
involved is a left-mover, only the transverse components
$\vec{A}_{\perp}$ of the right-mover give a dominant contribution to
the amplitude of the process\cite{Mueller:1999yb}. On the other
hand, in a fast moving bound state, since the two particles involved
are both, say, right-movers, it is $A^{-}$ that plays a more
important role since it apears in the Dirac equation via the
product with $p^+$ to contribute a term of order $\alpha^2$
while the contribution from $\vec{A}_\perp$ with the product of
$\vec{p}_\perp$ is of order $\alpha^3$.


\begin{thebibliography}{99}
\bibitem{Leitch:2008vw}
  M.~J.~Leitch  [PHENIX Collaboration],
  arXiv:0806.1244 [nucl-ex], references therein.

\bibitem{Matsui:1986dk}
  T.~Matsui and H.~Satz,
  Phys.\ Lett.\  B {\bf 178}, 416 (1986).

\bibitem{Liu:2006nn}
    K.~Peeters, J.~Sonnenschein and M.~Zamaklar,
  Phys.\ Rev.\  D {\bf 74}, 106008 (2006)
  [arXiv:hep-th/0606195]; H.~Liu, K.~Rajagopal and U.~A.~Wiedemann,
  Phys.\ Rev.\ Lett.\  {\bf 98}, 182301 (2007)
  [arXiv:hep-ph/0607062]; M.~Chernicoff, J.~A.~Garcia and A.~Guijosa,
  JHEP {\bf 0609}, 068 (2006)
  [arXiv:hep-th/0607089].

\bibitem{Adil:2006ra}
  A.~Adil and I.~Vitev,
  Phys.\ Lett.\  B {\bf 649}, 139 (2007)
  [arXiv:hep-ph/0611109].

\bibitem{Chu:1988wh}
  M.~C.~Chu and T.~Matsui,
  Phys.\ Rev.\  D {\bf 39}, 1892 (1989).

\bibitem{Peskin:1995ev}
  M.~E.~Peskin and D.~V.~Schroeder,
  An Introduction To Quantum Field Theory
(USA: Addison-Wesley, 1995).

\bibitem{Eichten:1978tg}
  E.~Eichten, K.~Gottfried, T.~Kinoshita, K.~D.~Lane and T.~M.~Yan,
  Phys.\ Rev.\  D {\bf 17}, 3090 (1978)
  [Erratum-ibid.\  D {\bf 21}, 313 (1980)].

\bibitem{Feynman:1949hz}
  R.~P.~Feynman,
  Phys.\ Rev.\  {\bf 76}, 749 (1949).

\bibitem{Kogut:1969xa}
  J.~B.~Kogut and D.~E.~Soper,
  Phys.\ Rev.\  D {\bf 1}, 2901 (1970).

\bibitem{Kapusta:1989tk}
  J.~I.~Kapusta,
  Finte Temperature Field Theory(Cambridge University Press, 1989).

\bibitem{LeBellac:1991cq}
  M.~Le Bellac,
  Thermal Field Theory(Cambridge University Press, 1996).

\bibitem{Kobes:1990dc}
  R.~Kobes, G.~Kunstatter and A.~Rebhan,
  Nucl.\ Phys.\  B {\bf 355}, 1 (1991).

\bibitem{Mueller:1999wm}
  A.~H.~Mueller,
  Nucl.\ Phys.\  B {\bf 558}, 285 (1999)
  [arXiv:hep-ph/9904404].

\bibitem{Mueller:1999yb}
  A.~H.~Mueller,
  arXiv:hep-ph/9911289;  A.~H.~Mueller,
  arXiv:hep-ph/0111244.

\bibitem{MultipleScattering}
 Fabio~Dominguez, C.~Marquet and Bin~Wu,
 in preparation.

\bibitem{Baier:1996kr}
  R.~Baier, Y.~L.~Dokshitzer, A.~H.~Mueller, S.~Peigne and D.~Schiff,
  Nucl.\ Phys.\  B {\bf 483}, 291 (1997)
  [arXiv:hep-ph/9607355].

\bibitem{Arnold:2000dr}
  P.~Arnold, G.~D.~Moore and L.~G.~Yaffe,
  JHEP {\bf 0011}, 001 (2000)
  [arXiv:hep-ph/0010177].

\bibitem{Moore:2004tg}
  G.~D.~Moore and D.~Teaney,
  Phys.\ Rev.\  C {\bf 71}, 064904 (2005)
  [arXiv:hep-ph/0412346].

\bibitem{Laine:2008cf}
  M.~Laine,
  arXiv:0810.1112 [hep-ph].

\bibitem{Escobedo:2008sy}
  M.~A.~Escobedo and J.~Soto,
  arXiv:0804.0691 [hep-ph].


\end{thebibliography}
\end{document}